\documentclass{ws-procs9x6}            

\begin{document}

\title{Wave-Selection Techniques for Partial-Wave
Analysis in Light-Meson Spectroscopy}

\author{F. M. Kaspar$^*$}
\author{B. Grube, F. Krinner$^\dagger$, S. Paul, S. Wallner}
\address{\scriptsize Physics Department, Technical University of Munich,\\
Garching bei M\"unchen, Bavaria 85748, Germany\\
$^\dagger$Max Planck Institut for Physics,\\
80805 Munich, Bavaria, Germany\\
$^*$E-mail: florian.kaspar@tum.de\\}

\begin{abstract}
The light-meson spectrum can be studied by analyzing data from
diffractive dissociation of pion or kaon beams. 
The contributions of the
various states that are produced in these reactions are disentangled
by the means of partial-wave analysis. A challenge in these analyses is
that the partial-wave expansion has to be truncated, i.e. that only a
finite subset of the infinitely many partial-wave
amplitudes can be inferred from the data. In recent years, different
groups have applied regularization techniques in order to determine the
contributing waves from the data. However, to obtain meaningful results
the choice of the regularization term is crucial. We present our recent developments of
wave-selection methods for partial-wave analyses based on simulated
data for diffractively produced three-pion events.
\end{abstract}

\small{}

\bodymatter
\section{Introduction}\label{intro:sec1}
The light-meson spectrum can be studied by analyzing data from
diffractive dissociation of pion or kaon beams.
The intensity distribution of the particles is usually modeled by the isobar model and the partial-wave analysis is performed using log-likelihood fits in bins of the final-state mass. A detailed description can be found in Ref.~\citenum{PhysRevD.95.032004}.
The model parameters are a set of complex numbers $\{T_i\}$, each corresponding to the transition amplitude of a partial wave $i$. The number of partial waves in the decomposition is in principle infinite. However, only a finite subset can be fitted to data.
Historically, this subset has been selected manually. However, this procedure may introduce hard-to-diagnose biases. It is therefore desirable to apply an algorithm to select relevant waves.
In Ref.~\citenum{Guegan:2015mea}, the selection was performed using the so-called LASSO method, described in Ref.~\citenum{tibshirani}, which has also been applied to Baryon Spectroscopy, e.g., in Ref.~\citenum{Landay2017}. The method regularizes the fit to the data by adding a penalty term to the log-likelihood function and thereby forces the intensities $\{|T_i|^2\}$ of insignificant waves to zero in order to exclude them from the model. In Ref.~\citenum{Bicker2016}, a similar method has been used, albeit with a different penalty term.
We compare both penalty terms on simulated data for the process $\pi^- + p \rightarrow \pi^-\pi^-\pi^+ + p$ and suggest extensions that combine the advantages of both. We address strategies for tuning free parameters and the issue of multiple fit solutions.
\section{Model Selection}
We demonstrate the effects of different regularization procedures on a Monte Carlo data set of approximately $37\times10^3$ events in the region 1.80\,GeV~$<~m_{3\pi}~<$~1.82\,GeV.
The data have been generated according to a reference wave set of 126 waves. The intensities of the individual waves vary over several orders of magnitude in order to imitate real data.

\begin{figure}[!tbp]
	\begin{minipage}[b]{0.49\textwidth}
		\includegraphics[width=\textwidth]{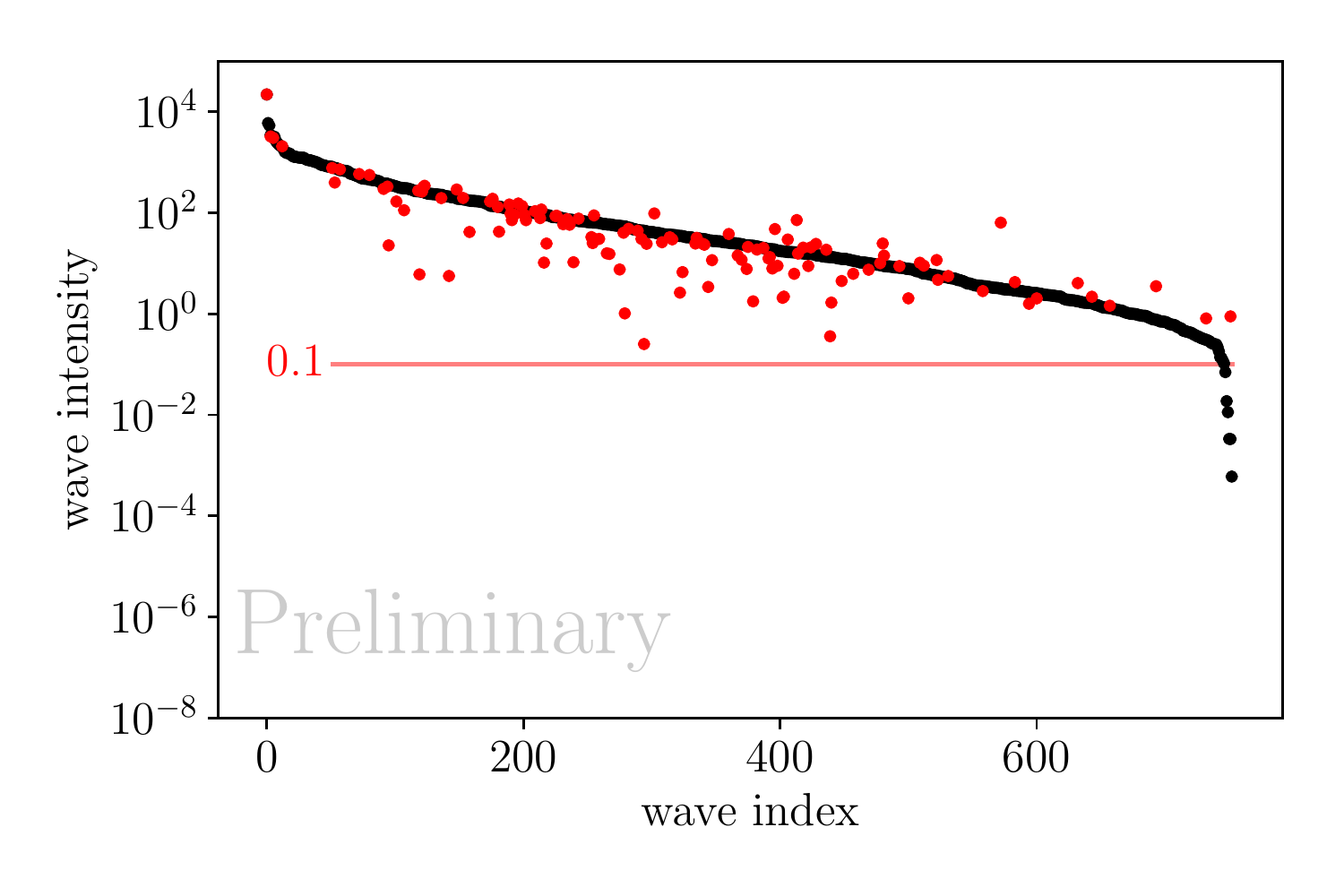}
		\vspace*{-10mm}
		\caption{Fit without regularization.}
		\label{aba:noreg}
	\end{minipage}
	\begin{minipage}[b]{0.49\textwidth}
		\includegraphics[width=\textwidth]{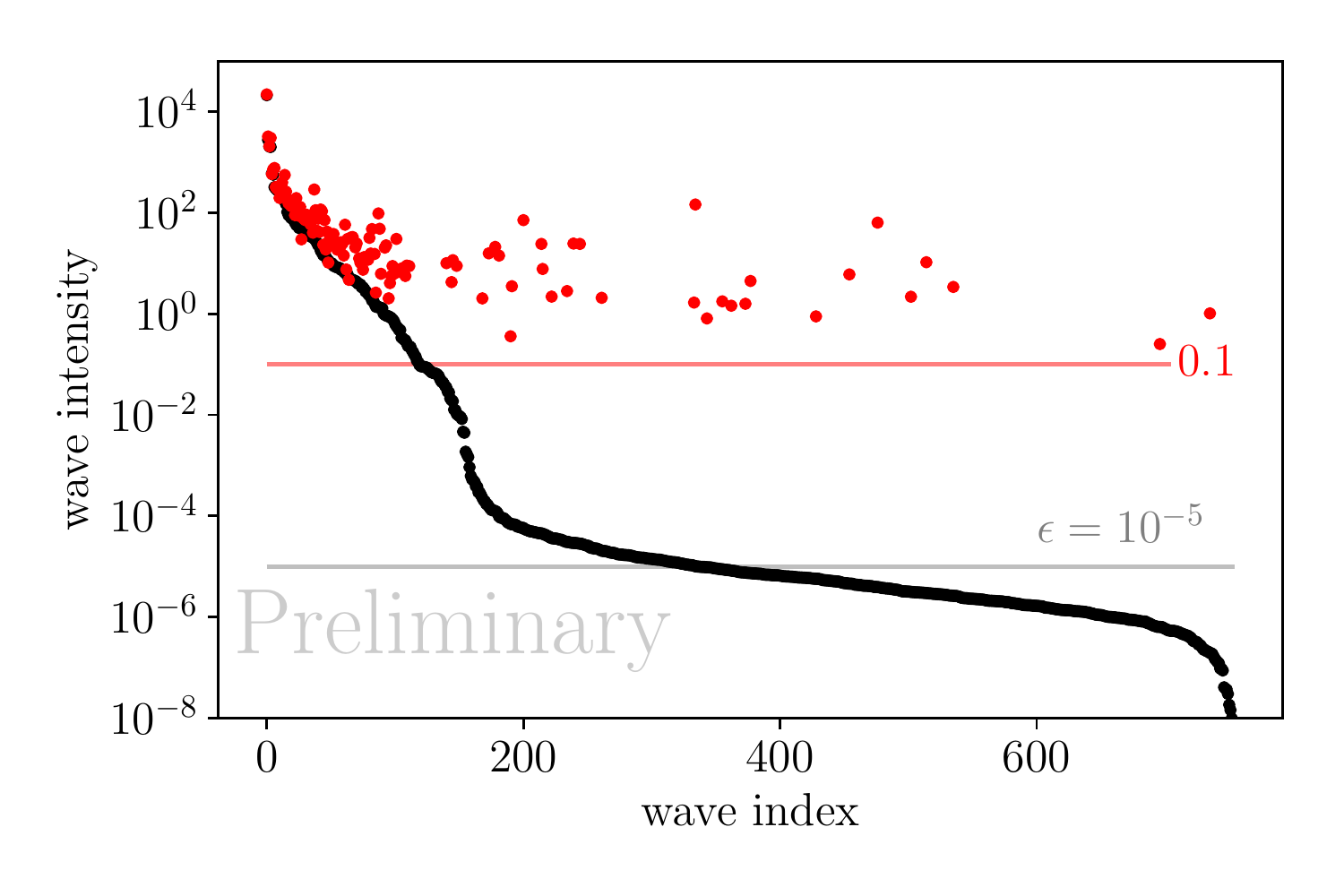}
		\vspace*{-10mm}
		\caption{Fit with LASSO regularization.}
		\label{aba:lasso}
	\end{minipage}
	\begin{minipage}[b]{0.49\textwidth}
		\includegraphics[width=\textwidth]{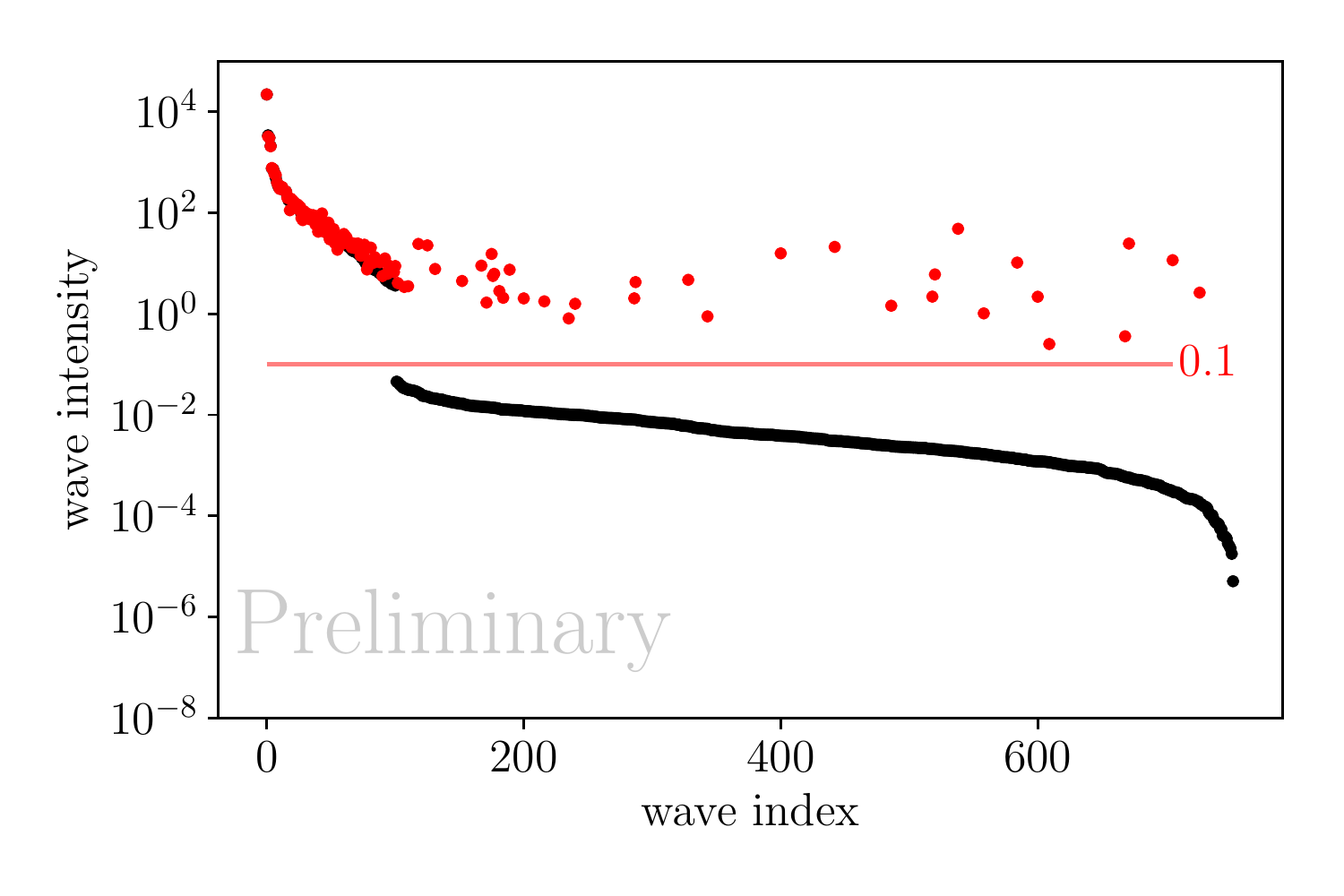}
		\vspace*{-10mm}
		\caption{Fit with BCM regularization.}
		\label{aba:bcm}
	\end{minipage}
	\begin{minipage}[b]{0.49\textwidth}
		\includegraphics[width=\textwidth]{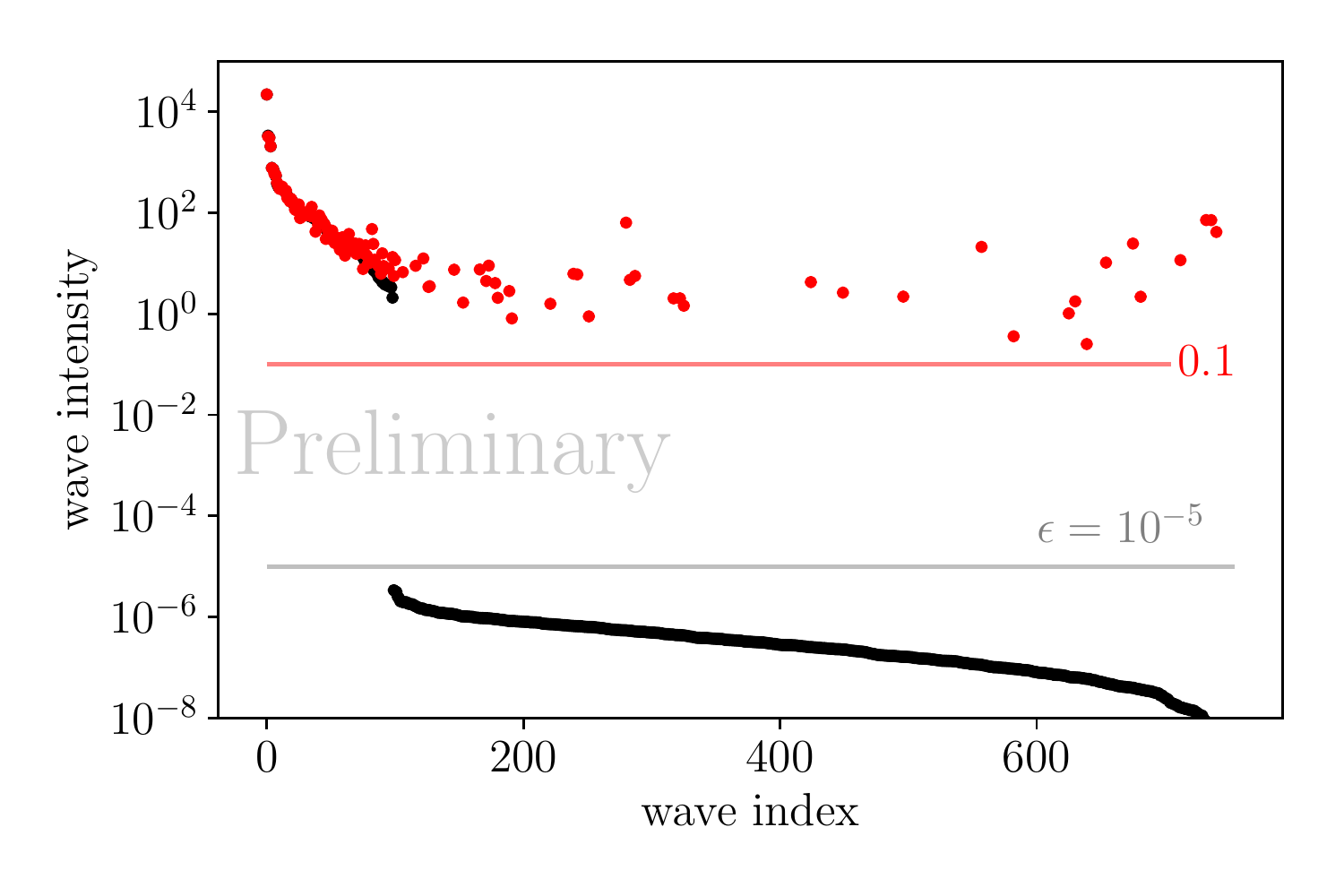}
		\vspace*{-10mm}
		\caption{Fit gen. Pareto regularization.$^\text{a}$}
		\label{aba:pareto}
	\end{minipage}
\end{figure}

The selection procedure starts from a large, systematically constructed wave pool where the spin and angular momentum quantum numbers are limited. In this case, there are 753 waves in the pool. An unregularized fit of the whole wave pool is not able to reproduce the reference model. \Fref{aba:noreg} shows the partial-wave intensities of the fitted model (black) in descending order with the reference fit intensities (red) sorted accordingly. For many waves,  the intensities of the fitted waves do not match the ones of the reference model.
\Fref{aba:lasso} demonstrates the effect of the LASSO penalty, i.e.
\begin{equation}
	\log{\mathcal{L}} - \frac{1}{\Gamma}\sum_{i}{\left| T_{i} \right|},
\end{equation}
where $\Gamma=0.3$ in this case. See section \ref{sec:tuning} for our plans on parameter tuning.

To avoid discontinuities of the derivatives of $\mathcal{L}$ close to $T_i=0$, we use the approximation $|T_i|\approx\sqrt{|T_i|^2+\epsilon}$ with $\epsilon=10^{-5}$. We observe a smooth drop in intensity down to the scale set by $\epsilon$. For most large-intensity waves, we correctly recover the reference values. However, we observe a bias towards smaller intensities in the regularized fit.

\Fref{aba:bcm} demonstrates the effect of the  ``Biggest Conceivable Model'' method (BCM), that has been developed in Ref.~\citenum{Bicker2016} and has first been applied to $3\pi$ real data in Ref.~\citenum{Drotleff}. The penalty term is the logarithm of a Cauchy distribution, i.e.
\begin{equation}
	\log{\mathcal{L}} - \sum_{i}{\log{ \left(  1+\left| T_{i} \right|^2/\Gamma^2\right) }},
\end{equation}
where $\Gamma=0.2$ for \fref{aba:bcm}. The BCM method yields a discontinuous drop in intensity indicating the selection of waves above the drop. This behavior appears for sufficiently small value of $\Gamma$. In contrast to the LASSO, the bias towards smaller intensities is significantly reduced. This is because the penalty term rises only logarithmically. However, the deselected waves are not pushed all the way to zero (or $\epsilon$) intensity, which may lead to a bias.

We also investigate the usefulness of other penalty terms that combine advantages of both of the aforementioned ones.
In Ref.~\citenum{Armagan2013}, the generalized Pareto distribution, i.e.
\begin{equation}
	\log{\mathcal{L}} - \frac{1}{\zeta}\sum_{i}{\log{ \left(  1+\zeta\left| T_{i} \right|/\Gamma \right) }},
\end{equation}
has been proposed for selection. Its behavior for $\Gamma = 0.1$ and $\zeta = 0.5$ shown in \Fref{aba:pareto} is very similar to the BCM method for large intensities and to the LASSO for small intensities. The deselected waves have almost zero intensity, which further reduces their influence on the selected waves. For $\zeta\rightarrow0$ the penalty approximates the LASSO.

Due to destructive interference, partial waves may not contribute significantly to the total intensity although being very large. A logarithmic penalty becomes too flat in order to provide a useful regularization in this case. Introducing a linear rise of the penalty by adding an appropriate term similar to the smoothed approximation of the LASSO (but with larger effective $\epsilon$), i.e.
\begin{equation}
	\log{\mathcal{L}} - \sum_{i}{\left[\log{\left({ 1+\left|T_i\right|^2/\Gamma^2 }\right)} +  \frac{1}{\zeta} \sqrt{ \left|T_i\right|^2/\Gamma^2+1 } -  \frac{1}{\zeta}\right]} ,
\end{equation}
can provide a trade-off between bias on the intensity and the introduction of destructive interference. The second free parameter $\zeta$ adjusts the relative strengths of the two terms.

\begin{figure}[!tbp]
	\centering
	\begin{minipage}[b]{0.47\textwidth}
		\includegraphics[width=\textwidth]{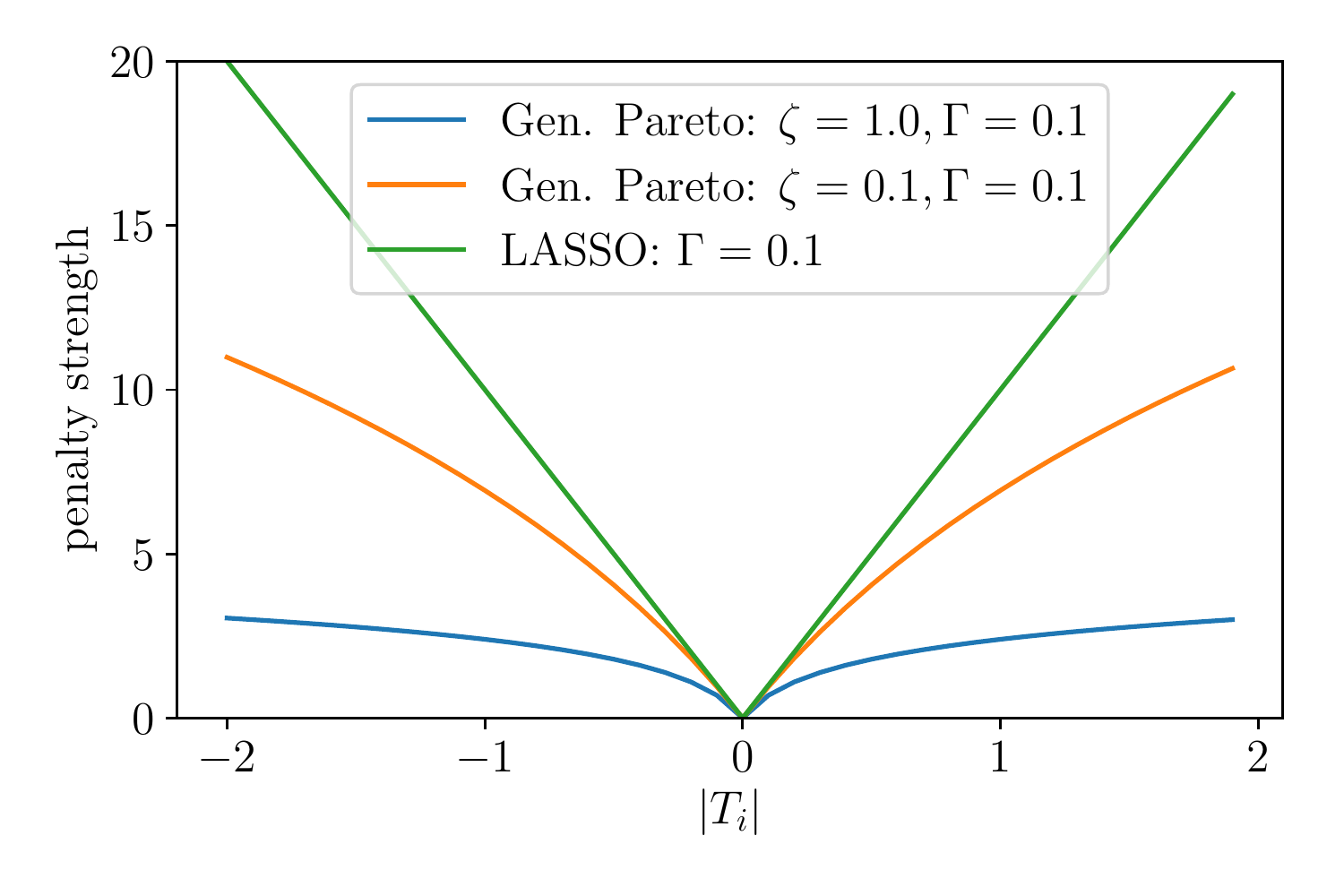}
		\vspace*{-10mm}
		\caption{Comparison of the LASSO with generalized Pareto.}
		\label{aba:compareLassoLike}
	\end{minipage}
	\hfill
	\begin{minipage}[b]{0.47\textwidth}
		\includegraphics[width=\textwidth]{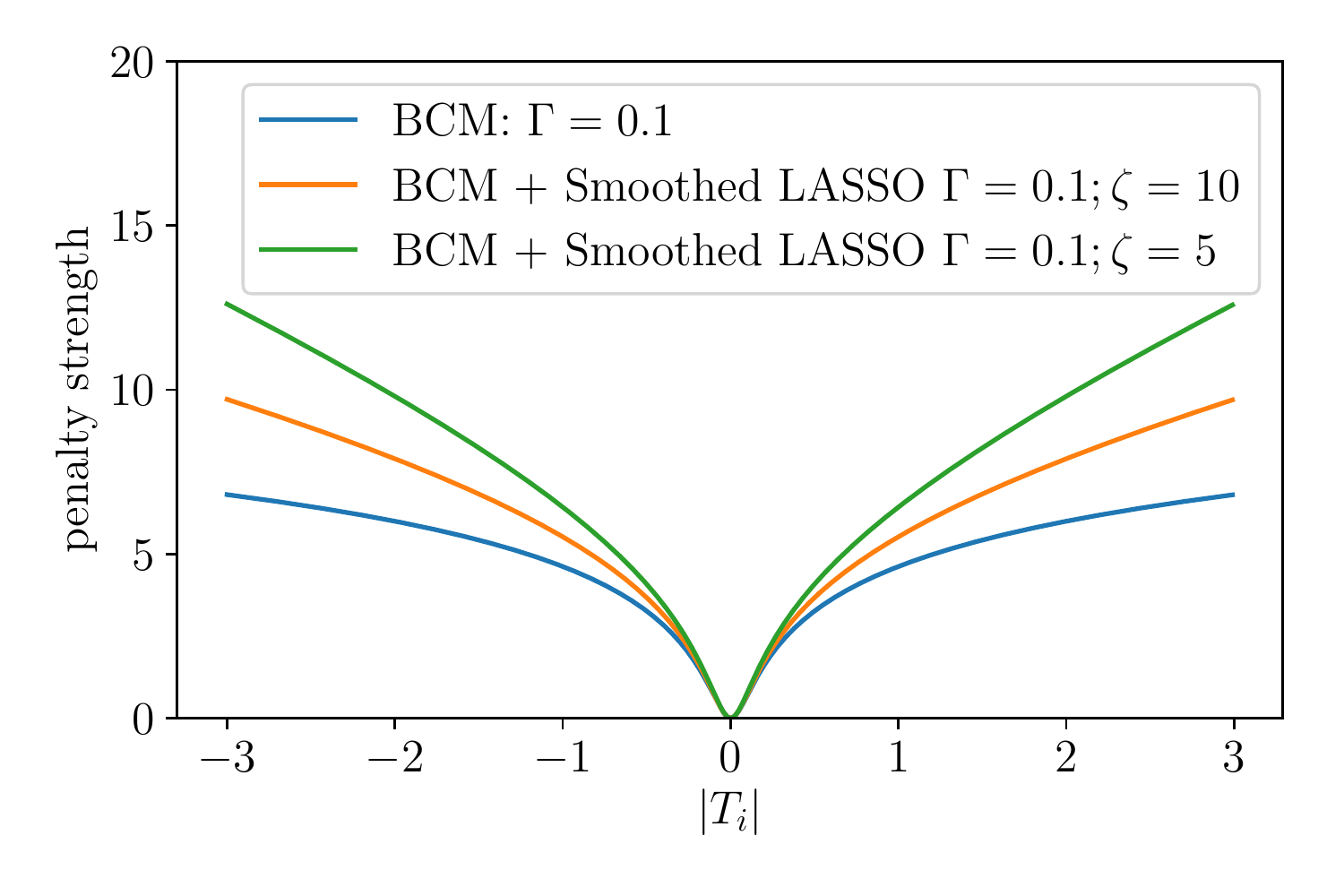}
		\vspace*{-10mm}
		\caption{Comparison of BCM with and without additional LASSO term.}
		\label{aba:compareCauchyLike}
	\end{minipage}
\end{figure}

A comparison of the different penalties is shown in \fref{aba:compareLassoLike} and \fref{aba:compareCauchyLike}.
\footnotetext{$^\text{a}$~This plot has been added after the conference to demonstrate this penalty on the same data set.}
\section{Parameter Tuning}
\label{sec:tuning}
All penalty terms have free parameters and, a priori, it is not clear how to choose them. If the regularization is too strong, the fit will not describe the data, if it is too weak, the fit will describe fluctuations and a similar situation like in \fref{aba:noreg} will arise. In Ref.~\citenum{Guegan:2015mea}, the authors rely on the information criteria AIC\cite{1100705} and BIC\cite{schwarz1978}. However, it is not clear whether the assumptions required for their applicability are fulfilled in our case (see references for details). Therefore, we are currently investigating a cross-validation procedure similar to Ref.~\citenum{Landay2017}.
\section{Coping with Multiple Fit Solutions}
\begin{figure}[!tbp]
	\centering
	\begin{minipage}[b]{0.47\textwidth}
		\includegraphics[width=\textwidth]{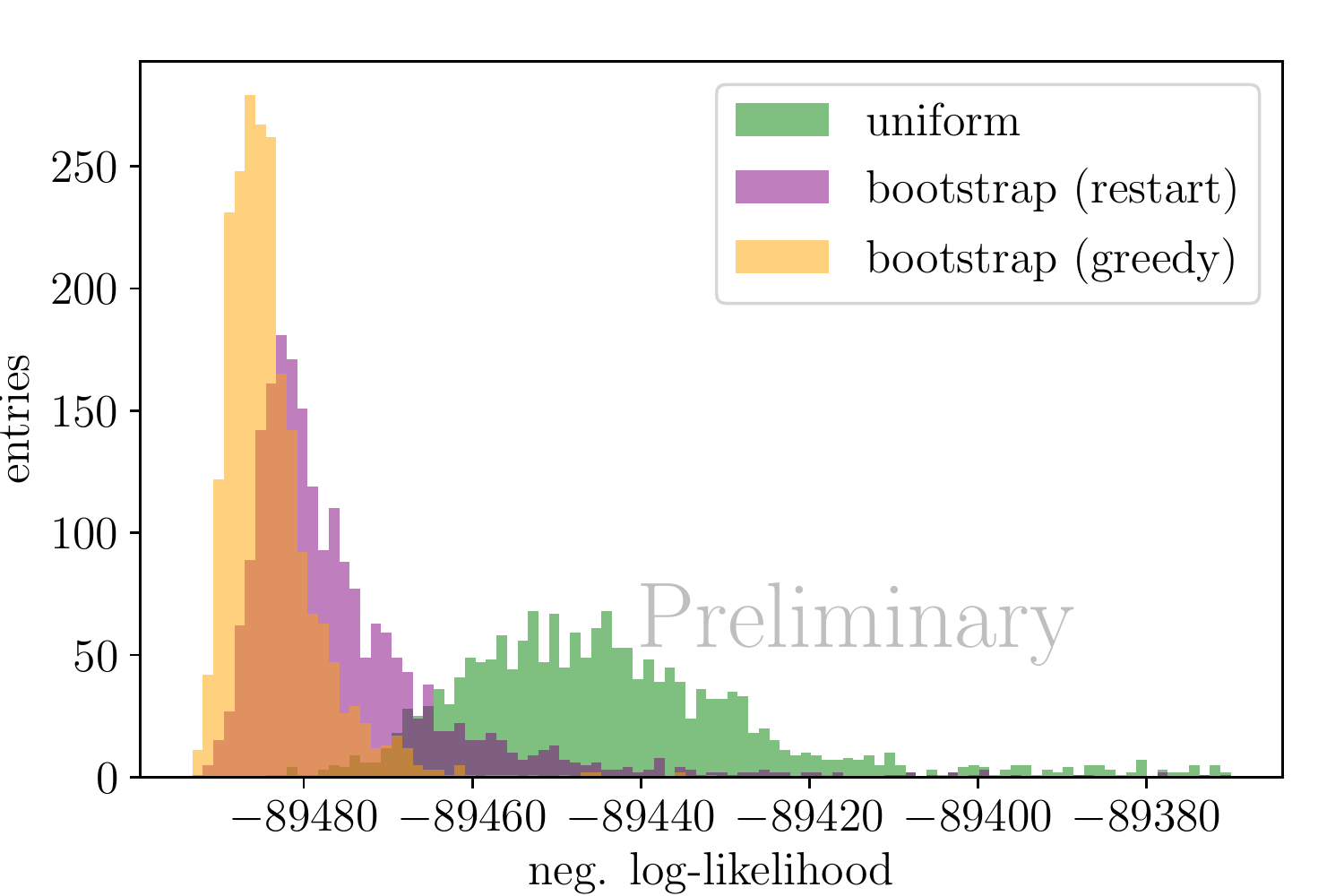}
		\vspace*{-5mm}
		\caption{Distribution of negative log-likelihood values for different start-parameter generation methods.}
		\label{aba:histo}
	\end{minipage}
\hfill
	\begin{minipage}[b]{0.47\textwidth}
		\includegraphics[width=\textwidth]{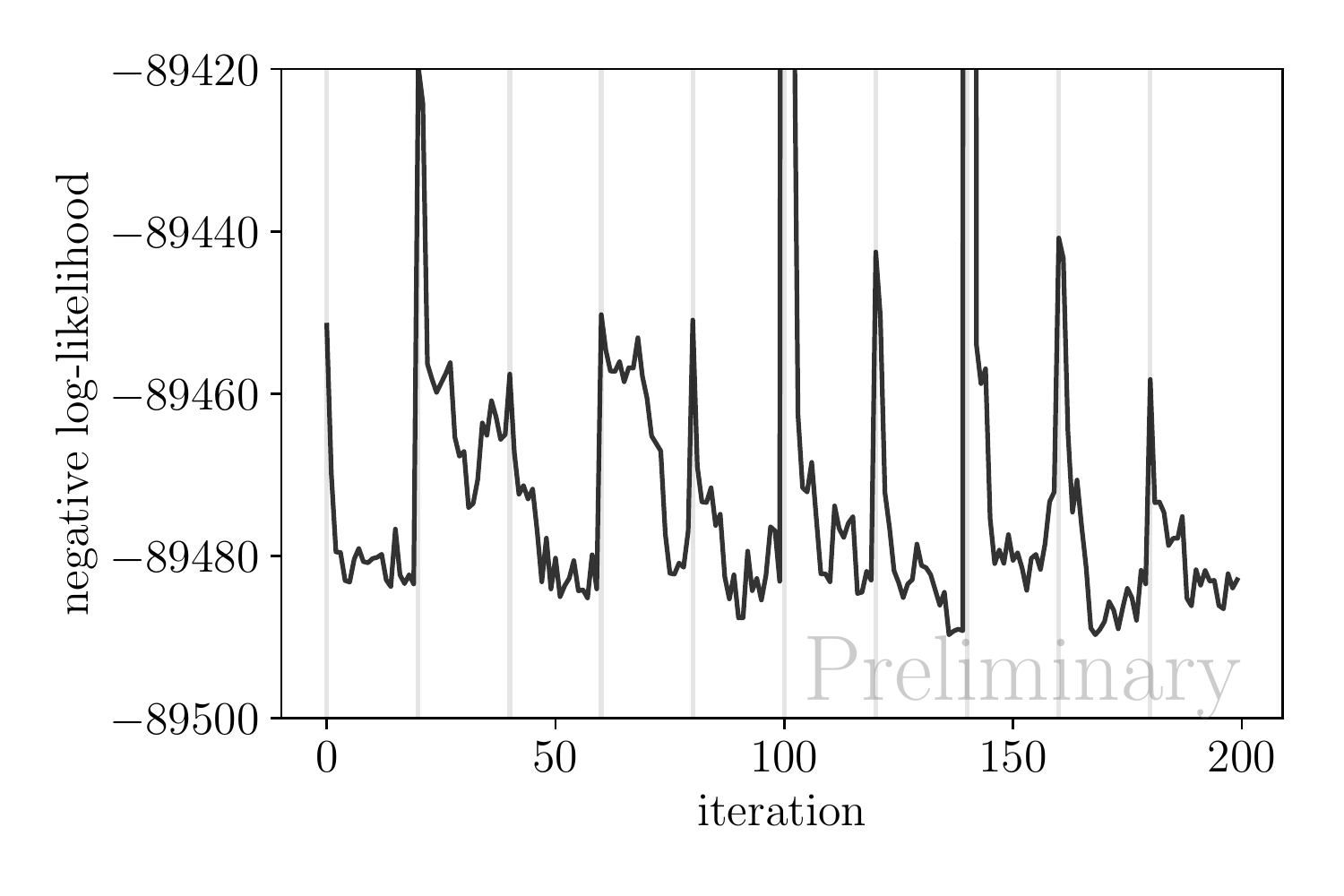}
		\vspace*{-5mm}
		\caption{Negative log-likelihood for boostrap start-parameter generation. Gray lines mark hard restarts.}
		\label{aba:improv}
	\end{minipage}
\end{figure}

The model-selection fits suffer from many local minima of the negative log-likelihood function. This causes a strong dependence of the fit result on the start parameter values. For every fit, we sample multiple random start values and perform the optimization for each of them. We use the parameters with the smallest negative log-likelihood as our best estimate.
We investigate the bootstrap restarting method described in Ref.~\citenum{Wood2001} as a more efficient alternative to uniform start parameter sampling. The method uses parameter estimates of fits on a perturbed data set as a start-parameter proposal. We also consider a mixed approach, where a uniform draw of the start parameters is performed on every 20th step (hard restart) as opposed to only using the bootstrap ansatz (greedy). \Fref{aba:histo} shows the concentration of the negative log-likelihood towards smaller values for both bootstrap methods in contrast to the spread out distribution for uniform sampling. In \Fref{aba:improv} the value of the negative log-likelihood is shown over the number of draws for the reset method. The uniform restarts are marked by a light gray line every 20th step. How well this heuristic approach performs on our problem in practice is currently under investigation. 

\small
\section*{Acknowledgment}\label{outl:acknl}
This work was supported by the BMBF, the Excellence Cluster ORIGINS which is funded by the Deutsche Forschungsgemeinschaft (DFG, German Research Foundation) under Germany's Excellence Strategy – EXC-2094 – 390783311, and the
Maier-Leibnitz-Laboratorium der Universität und der Technischen
Universit\"at M\"unchen.

\bibliographystyle{ws-procs9x6}
\bibliography{Biblio/bibliography}
\end{document}